\documentclass[twocolumn,showpacs,preprintnumbers,amsmath,amssymb]{revtex4}
\renewcommand{\O}{\Omega}
\renewcommand{\o}{\omega}
\newcommand{\ti}{\tilde}
\renewcommand{\sc}{\mathscr{I}^+}
\newcommand{\scc}{\mathscr{I}^-}

\newcommand{\ri}{\rightarrow}
\newcommand{\bq}{\begin{equation}}
\newcommand{\ee}{\end{equation}}
\newcommand{\f}{\frac}

\newcommand{\bee}{\begin{equation}}
\newcommand{\eee}{\end{equation}}
\newcommand{\beq}{\begin{eqnarray}}
\newcommand{\eeq}{\end{eqnarray}}
\newcommand{\na}{\tilde{\nabla}}
\newcommand{\pa}{\partial}
\newcommand{\sq}{\sqrt}

\newcommand{\bc}{\begin{center}}
\newcommand{\ec}{\end{center}}
\newcommand{\bl}{\biggl}
\newcommand{\br}{\biggr}
\newcommand{\beqq}{\begin{eqnarray*}}

\newcommand{\no}{\nonumber}

\usepackage{graphicx}
\usepackage{epsbox}
\usepackage{ascmac}
\usepackage{rsfs}

\begin{document}

\title{Topological Structure of The Upper End of Future Null Infinity}

\author{Shinya Tomizawa}
\email{tomizawa@th.phys.titech.ac.jp}
\affiliation{Department of Physics, Tokyo Institute of Technology, Oh-Okayama, Tokyo 152-8550, Japan }

\author{Masaru Siino}
\email{msiino@th.phys.titech.ac.jp}
\affiliation{Department of Physics, Tokyo Institute of Technology, Oh-Okayama, Tokyo 152-8550, Japan}
\date{\today}

\begin{abstract}
 We propose a method to determine the topological structure of an event horizon far in the future of a spacetime from the geometrical information of its future null infinity. In the present article, we mainly consider spacetimes with two black holes. Although, in most of cases, the black holes coalesce and their event horizon is topologically a single sphere far in the future, there are several possibilities that the black holes do not coalesce eternally and such exact solutions. 

In our formulation, the geometrical structure of future null infinity is related to the topological structure of the upper end of the future null infinity through the Poincare\'-Hopf's theorem. Since the upper end of the future null infinity determines the event horizon far in the future under the conformal embedding, the topology of event horizon far in the future will be affected by the geometrical structure of the future null infinity. Our method is not only for the case of black hole coalescence. Also we can consider more than two black holes or a black hole with non-trivial topology.
\end{abstract}


\maketitle

\section{\label{sec:1}Introduction}
We are often interested in the final state of a black hole spacetime after gravitational collapse. When matter collapses, it passes inside an event horizon and energy will be emitted to future null infinity in the form of gravitational waves. Since the amount of this energy is limited, we may expect that the spacetime will approach to a stationary state. If a black hole spacetime becomes stationary far in the future, the final state of the black holes is considerably restricted.

So far the final state of the black hole has been studied by many authors and these studies are known as uniqueness theorem of black holes. Israel showed that the only static and topologically spherical black hole is Schwartzshild solution \cite{Israel} or Reissner-Nortstr\"{o}m solution \cite{Israel2}. In stationary axisymmetric situations, the uniqueness theorems of a vacuum black hole and a charged black hole were shown by Carter \cite{Carter} and Robinson \cite{Robinson},\cite{Robinson2}. Hawking proved that a stationary black hole have to be static or axisymmetric and that the spatial topology of the horizon is the only sphere \cite{Hawking}.

In these works, however, the existence of endpoints of event horizon generators was not considered. In dynamical situation, the endpoints of event horizon generators, where the event horizon is indifferentiable, are very important when we discuss the topology of an event horizon, because it causes the change of the spatial topology of an event horizon \cite{Siino}. This is revealed by one of authors \cite{Siino}. When this was discussed in \cite{Siino}, the topology of an event horizon far in the future is assumed to be a sphere. On the other hand, the work of Chru\'sciel and Wald~\cite{Chrusciel and Wald} implies that in a stationary spacetime, regardless of the existence of a endpoint, the spatial topologies of the connected components of black holes are only spheres under null energy condition.
Therefore, if far in the future, the black hole spacetime settles down to a stationary state, we expect the spatial topology of the event horizon will consist of some connected components with spherical topology. To realize the relation of these works, it is required to determine the topology of an event horizon far in the future. There is not the simple method to know topological structure of an event horizon far in the future, that is, to determine how many black holes the final state of the black hole spacetime will consist of. We will develop such a method in the present article. 

Since in an asymptotically flat spacetime \cite{Wald}, (to be more exact, in a strongly asymptotically predictable spacetime \cite{Wald} ) an event horizon is defined as the boundary of the causal past of future null infinity, we expect the topology of an event horizon will be related to the topological structure of future null infinity. Because the topology of future null infinity in an asymptotically flat spacetime is $S^2\times {\bf R}$, we may conclude far in the future the topology of an event horizon will always be a single $S^2$. Nevertheless it is well known that there are some stationary solutions including black holes which do not coalesce eternally. For example, they are Majumdar-Papapetrou solution \cite{Hartle and Hawking}, C-metric \cite{Pravda} and so on. The spatial topology of the event horizon of such a solution is obviously not a single sphere. In these spacetimes, is the topology of the upper end of future null infinity really $S^2$ ? (Of course, since the future null infinity is open at the upper end, this statement may be not well-defined.) If we regard that the topology of the upper end of future null infinity determines the topology of the event horizon far in the future, we may say that the topology of the upper end of future null infinity in such spacetimes is not one sphere but two spheres.

In the present article, we consider that in such spacetimes with the black holes which do not merge eternally, the topological structure at the upper end of future null infinity pinches. Here we emphasize that its topological structure determines the spatial topology of an event horizon far in the future. In section \ref{sec:2}, we will give the index theorem and mention the relation between the topology of the upper end of future null infinity and geometrical structure of the null congruence of the generator on future null infinity. Since we discuss black hole spacetimes, we consider the only asymptotically flat spacetime. So we will give its definition in the rest of section \ref{sec:2}. In section \ref{sec:4}, we will discuss the method to examine the topology of the upper end of future null infinity. In section \ref{sec:5}, we will apply this method to C-metric as an example, to illustrate the detail concretely.  

Throughout this article, we use the abstract index notation as the component notation of tensors and it is denoted by Latin indices $a,b,\cdots$.

\section{\label{sec:2}Preparation }
How can we know the topology of an event horizon far in the future?? As mentioned above, an event horizon is defined as the boundary of the causal past of future null infinity, whose topology is $S^2\times{\bf R}$ \cite{Wald}. So at first sight we think that the spatial topology of black holes is a single sphere finally. That is, all black holes coalesce and become a single far in the future. Is this real ? There are some exceptions to this. The examples of these exceptions are Majumdar-Papapetrou solution and C-metric. On the former, each black hole does not coalesce eternally by the force of electric repulsion. The latter describes two uniformly accelerated black holes connected by a string which are away from each other in the future. It is evident that in these spacetime black holes do not coalesce far in the future. It seems that these are inconsistent with the topology of future null infinity, $S^2\times{\bf R}$. 

Nevertheless, there is doubt that the topology of the upper end of $\sc$ is a sphere (FIG.1(a)). For because the definition of $\mathscr{I}^+$ is $\sc\equiv\dot{J}^+(i^0)-i^0$, it is open to the future, where $J^+(i^0)$ is the causal future of spatial infinity $i^0$. Therefore, there is not evidence that the topology of the upper end of $\mathscr{I}^+$ is a sphere in such spacetimes. As shown in FIG.1(b), we may expect that $\sc$ pinches at the upper end in the spacetime where two black holes does not eternally coalesce. 

\begin{figure}[htbp]
\begin{center}
\includegraphics[width=.70\linewidth]{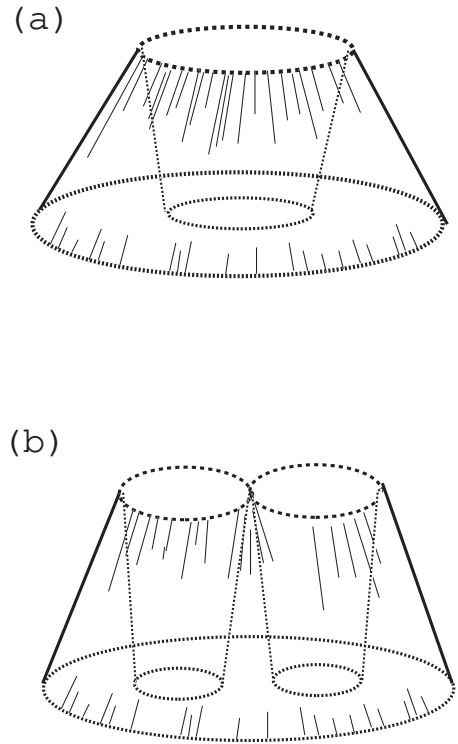}
\end{center}
\caption{The topological structure of the upper end of future null infinity in the case (a) where two black holes will coalesce in the future and in the case (b) where two black holes will not collide far in the future. In (a), $\sc$ is a sphere at the upper end but in (b), $\sc$ pinches at the upper end. }
\end{figure}

First of all, we must mention the method to examine the topology of the upper end of future null infinity. The following corollary \cite{Sorkin} of the Poincar\'e-Hopf's theorem is useful.

\newtheorem{theo}{Theorem}

\begin{theo}
Let M be a compact n-dimensional ($n>2$ is an odd number) $C^r(r\ge 1)$ manifold with $\Sigma_1\cup\Sigma_2=\partial M$ and $\Sigma_1\cap\Sigma_2=\emptyset$. X is any $C^{r-1}$ vector field with at most a finite number of zeros, satisfying the following two conditions: (a)The zeros of X are contained in Int M. (b)X has inward directions at $\Sigma_1$ and outward directions at $\Sigma_2$. Then the sum of the indices of X at all its zeros is related to the Euler numbers of $\Sigma_1$ and $\Sigma_2$:
\begin{equation}
\chi(\Sigma_2)-\chi(\Sigma_1)=2\ {\rm index}(X), \label{eq:theorem}
\end{equation}
\end{theo}    
where index$(X)$ is given by the alternating sum of the Morse number $\mu_k$ as $\rm{index}=\sum_k (-1)^k\mu_k$. The Morse number $\mu_k$ is the number of a critical point (zero of vector field $X$) whose index is $k$. The index of a critical point is given by the number of negative eigenvalue of Hesse matrix $H_{ab}=\nabla_a X_b$.  

This theorem means that when the topology of slices of a manifold changes, there must be a zero of the vector field on it and equation (\ref{eq:theorem}) is satisfied.  Because $\sc$ is a three-dimensional null hypersurface, we can apply this theorem to $\sc$. Let us consider the tangent vector field of the null generators on $\sc$ as the vector field in this theorem. Here, we should pay attention to the fact that, in this case, zero can exist only on the upper end of $\sc$, since $\sc$ is geodesically complete about own null generators by definition. Therefore we have to be careful to apply the theorem. Nevertheless, it would be possible to relate the theorem of index and the congruence of null generators around zero. In FIG.2 and the following, we temporarily consider small extension of future null infinity into its future directions for an explanation. From Morse's lemma, we suppose that the zero of the null generators is isolated. Since $\sc$ is three dimension, the behavior of the congruence around zero can be classified and their index is determined. These are illustrated in FIG.2. The left figures are three dimensional manifolds with index$=\pm 1$ zeros, and the right represent the behavior of the null congruence on the neighborhood around their zeros. From this figure, there will be type (a) zero on upper end of $\sc$ with $\rm{index}=+1$. Without reaching to the upper end, the asymptotic behavior of shear and expansion tells us information about the zero. Then we expect shear dominates the congruence that is approaching to the type (a) zero, and expansion to the type (b)-(c) zero with $\rm{index}=-1$. If we assume that when two black holes do not coalesce, there should be a zero with $\rm{index} +1$ and we can say that the ratio of the shear to the expansion will diverge near the zero.

\begin{figure}[htbp]
\begin{center}
\includegraphics[width=.70\linewidth]{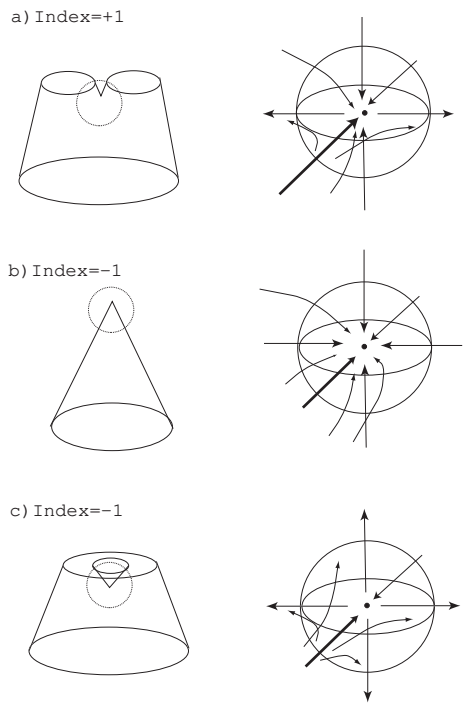}
\end{center}
\caption{The behaviour of vector field around the zeros of (a) index$=+1$ and (b),(c)index$=-1$. We can expect that in (a) the shear will become larger than the expansion and that in (b),(c) the shear will smaller than the expansion.}
\label{fig:2}
\end{figure}

Asymptotic flatness is essential in our discussion.
 We have to state the definition of asymptotically flat spacetime before explaining the analytic method to examine the topological structure of $\mathscr{I}^+$.  The definition mentioned here is based on the reference \cite{Wald}.

\newtheorem{df}{definition}
\begin{df}    
A spacetime $(M,g_{ab})$ is said to be asymptotically flat at null infinity if there exists an unphysical spacetime $(\tilde{M},\tilde{g}_{ab})$ with $\tilde{g}_{ab}=\O^2g_{ab}$ smooth everywhere except at the point $i^0$ where it is $C^{>0}$ and conformal isometry $\Psi:M\rightarrow\Psi (M)\in\tilde{M}$ with conformal factor $\Omega$ satisfying the below conditions. 
\begin{enumerate}
\item $\bar{J}^+(i^0)\cup\bar{J}^-(i^0)=\tilde{M}-M$. Thus, $i^0$ is spacelike related to all points in $M$ and the boundary of $M$ consists of $i^0$, $\sc\equiv\dot{J}^+(i^0)-i^0$ and $\scc\equiv\dot{J}^-(i^0)-i^0$.
\item There exists an open neighborhood $V$ of $\dot{M}=i^0\cup\sc\cup\scc$ such that the spacetime $(V,\tilde{g}_{ab})$ is strongly causal.
\item $\Omega$ can be extended to a function on all $\tilde{M}$ which is $C^{\infty}$ everywhere and  $C^{>0}$ at $i^0$.
\item On $\sc$ and $\scc$ we have $\O=0$ and $\na_a\O\ne 0$.
\item The map of null direction at $i^0$ into the space of integral curves of $n^a\equiv\tilde{g}^{ab}\tilde{\nabla}_b\Omega$ on $\sc$ and $\scc$ is diffeomorphism. 
\item For smooth function $\o$ on $\ti{M}-i^0$ with $\o>0$ on $\ti{M}\cup\sc\cup\scc$ which satisfies $\tilde{\nabla}_a(\omega^4n^a)=0$ on $\mathscr{I}^+$, the vector field $\omega^{-1}n^a$ is complete on $\sc$ and $\scc$.
\item In the neighborhood of $\sc$ and $\scc$ physical Ricci tensor behaviors as $R_{ab}=O(\Omega^2)$.
\end{enumerate}
\end{df}

Let us note that there is gauge freedom in the choice of an unphysical spacetime $(\tilde{M},\tilde{g}_{ab})$ with an asymptotically flat physical spacetime $(M,g_{ab})$~\cite{Wald}. This gauge freedom is most important in our discussion.
 
\textit{If the spacetime $(\tilde{M},\tilde{g}_{ab})$ is an unphysical spacetime satisfying the above definition with conformal factor $\O$, so is $(\tilde{M},\omega^2\tilde{g}_{ab})$ with conformal factor $\o\O$ for the any function $\omega$ which is smooth everywhere except at $i^0$ and positive everywhere.}

\section{\label{sec:4}analytic method }
In this section, we will investigate the relation of the geometrical structure of future null infinity (the congruence of the null generators and curvatures) and the topology of event horizon far in the future. 

\subsection{Physical Ricci Tensor and Unphysical Ricci Tensor}
The relation between the physical Ricci tensor and the unphysical Ricci tensor gives us a great deal of information. 
The physical Ricci tensor $R_{ab}$ is related to the conformally translated unphysical Ricci tensor $\tilde{R}_{ab}$ by

\beq
R_{ab}&=&\ti{R}_{ab}+2\O^{-1}\na_a\na_b\O \no\\
      &{}&+\ti{g}_{ab}\ti{g}^{cd}(\O^{-1}\na_c\na_d\O -3\O^{-2}\na_c\O\na_d\O).\label{eq:R}
\eeq

By multiplying $\O^2$ and taking the limit $\O\ri 0$ we find that a vector $n^a\equiv\ti{g}^{ab}\na_b\O$ must be extended smoothly to $\sc$ and be null at $\sc$ because the first term vanishes from conditions 4 and 7 of the definition 1, the second vanishes from conditions 3 and 4, and the fourth vanishes from the condition 3,4 and $\ti{g}_{ab}$ being smooth at $\sc$.

$n^a$ is null at $\sc$ but is not null off of $\sc$ and we put the shift from null vector as the following

\begin{equation}
\ti{g}_{ab}n^an^b=k_{(1)}\O+k_{(2)}\O^2+\cdots ,\label{eq:nn}
\end{equation}
where $k_{(i)}\ (i=1,2,\cdots)$ are the functions independent of $\O$. 
The integral curves of $n^a$ on $\sc$ is null geodesic generators of $\sc$, but the integral curves of $n^a$ is not geodesics off of $\sc$. Let us put the shift from the geodesic as
\begin{equation}
n^a\na_an^b-p_{(0)}n^b=O(\O).
\end{equation} 
In particular, we put the contraction of the above equation with $l^a\equiv (\pa/\pa\O)^a$ as the following:
\begin{equation}
l^bn^a\na_an_b=-p_{(0)}+p_{(1)}\O+\cdots .\label{eq:lnn}
\end{equation}
From (\ref{eq:nn}) and (\ref{eq:lnn}), we obtain the relation between $k_{(i)}$ and $p_{(i)}$
\begin{equation}
p_{(0)}=\f{1}{2}k_{(1)},\ p_{(1)}=-k_{(2)},\ p_{(2)}=-\f{3}{2}k_{(3)},\cdots ,\label{eq:pk}
\end{equation}
where we used the relation of torsion free ;
\begin{equation}
\na_an_b=\na_bn_a .
\end{equation}

\subsection{Semi-Newman-Penrose Formalism}
Newman-Penrose formalism~\cite{Newman Penrose} is the choice of the null basis which consists of a pair of real null vectors, $l^a, n^a$ and a pair of complex conjugate null vectors $m^a, \bar{m}^a$. Now we consider $n^a\equiv \ti{g}^{ab}\na_b\O$ as one of real null vectors in Newman-Penrose formalism. As we have mentioned before, $n^a$ is null on $\sc$ but is not null away off $\sc$. Therefore, strictly speaking, this is not Newman-Penrose formalism. Nevertheless the Newmann Penrose equations are not affected by this difference.

So the following orthogonality conditions are required
\beq
l^am_a=l^a\bar{m}_a=n^am_a=n^a\bar{m}_a=0\label{eq:orthogonal}
\eeq 
in addition to the conditions that vectors are null,

\begin{equation}
l^al_a=m^am_a=\bar{m}^a\bar{m}_a=0,
\end{equation}
and the fact that $n^a\equiv\ti{g}^{ab}\na_b\O$ is null only on $\sc$,

\begin{equation}
n^an_a=k_{(1)}\O+k_{(2)}\O^2+\cdots.\label{eq:nn2}
\end{equation}
We impose on the basis vectors the further normalization conditions,
\begin{equation}
l^an_a=-1,\ m^a\bar{m}_a=1,
\end{equation}
which is consistent with the definition of the vector $n^a$ and $l^a$.
Thus, the metric can be represented by 
\begin{equation}
\ti{g}_{ab}=(k_{(1)}\O+k_{(2)}\O^2+\cdots)l_al_b-2l_{(a}n_{b)}+2m_{(a}\bar{m}_{b)}\end{equation}
in the basis vectors. Note that the first extra term of the above equation is caused by the equation (\ref{eq:nn2}).

In order to investigate the behavior of the null geodesics on $\sc$, we have only to consider the values of spin connections and curvatures only in the neighborhood of $\sc$. 
Now we expand the each basis component of the equation (\ref{eq:R}) in the powers of conformal factor $\O$.  The $(n,l)$ component of the equation (\ref{eq:R}) gives
\beq
O(\O^2) &=& \bl(\f{5}{2}k_{(1)}-3p_{(0)}+2\mu_{(0)}\br)\O^{-1}\no\\
  &{}& +\bl(\ti{R}^{(0)}_{nl}+3p_{(1)}+2k_{(2)}+2\mu_{(1)}+k_{(1)}q\br)+O(\O),\no\\
\eeq
where we used the condition 7 of the above definition 1 and the equations (\ref{eq:nn}),(\ref{eq:lnn}) and $\mu\equiv-m^a\bar{m}^b\na_an_b$ is the spin connection, which means the expansion of the integral curves of $n^a$. $q$ is the function satisfying $l^a\na_al^b=ql^b$.

So we obtain from the coefficients of the first and the second term 
\begin{equation}
\f{5}{2}k_{(1)}-3p_{(0)}+2\mu_{(0)}=0\label{eq:mpk}
\end{equation}
and
\begin{equation}
\ti{R}^{(0)}_{nl}+3p_{(1)}+2k_{(2)}+2\mu_{(1)}+k_{(1)}q=0.
\end{equation}

Similarly, from other components we can obtain the relations between spin connections and Ricci tensors.

From the $(l,l)$ component, we obtain 
\begin{equation}
q=0,
\end{equation}
\begin{equation}
\ti{R}_{ll}^{(0)}=0.
\end{equation}
From the $(m,m)$ component, we obtain
\begin{equation}
\bar{\lambda}_{(0)}=\lambda_{(0)}=0,\label{eq:lam}
\end{equation}
\begin{equation}
\lambda_{(1)}=\ti{R}_{mm}^{(0)},
\end{equation}
where $\lambda\equiv -\bar{m}^a\bar{m}^b\na_an_b$ is the spin connection which means the shear of the integral curves of vector field $n^a$.
The $(n,m)$ component gives
\begin{equation}
\nu_{(0)}=\bar{\nu}_{(0)}=0,
\end{equation}
\begin{equation}
2\nu_{(1)}=\ti{R}_{nm}^{(0)},
\end{equation}
where $\nu$ is defined by $\nu\equiv n^a\bar{m}^b\na_an_b$. 

\subsection{Change of Conformal Factor}
Now it should be noted that there is gauge freedom in the conformal transformation considered above. Under the transformation $\O\ri\O'=\o\O, \ \ti{g}_{ab}\ri\ti{g}'_{ab}=\o^2\ti{g}_{ab}$, we have

\bq
n^a\ri n^{\prime a}=\o^{-1}n^a+\O\o^{-2}\ti{g}^{ab}\na_b\o.
\end{equation}
In particular, on $\sc$ the above transformation becomes
\begin{equation}
n^a\ri n^{\prime a}=\o^{-1}n^a.
\end{equation}

Under this transformation, the expansion of the null geodesic generators is transformed as
\begin{equation}
\mu_{(0)}\ri\mu'_{(0)}=\f{\mu_{(0)}}{\o}-\f{1}{\o^2}n^a\na_a\o\label{eq:mm}
\end{equation}
on $\sc$.
Therefore, note that we can choose $\mu_{(0)}$ as any function independent of $\O$ . Given $\mu'_{(0)}$, since  the equation (\ref{eq:mm}) is merely an ordinary differential equation, there always exists $\o$ satisfying the equation (\ref{eq:mm}). From the equations (\ref{eq:pk})(\ref{eq:mpk}), we find the relation
\begin{equation}
\mu_{(0)}=-p_{(0)}=-\f{1}{2}k_{(1)}.\label{eq:mupk}
\end{equation}
So from the equation (\ref{eq:lnn}), we see that in the gauge satisfying $\mu_{(0)}=0$, the null geodesics on $\sc$ are affinely parameterized. In this article, we call this gauge \textit{affine gauge}. When we discuss the upper end of $\sc$, we must not choose the affne gauge since in the affine gauge, there is not the upper end, that is, the null geodesics are complete in an affine parameterizing. Hence, we must choose the gauge such that $\sc$ is compact in the direction of $n^a$, that is, at the upper end $\o\ri 0$ along the null geodesics on $\sc$ . 

From (\ref{eq:lam}), the zeroth order for the shear $\lambda$, vanishes, which does not depend on the choice of the gauge; 
\begin{equation}
\lambda'_{(0)}=\lambda_{(0)}=0
\end{equation} 
holds.
Under the transformation $\O\ri\O'=\o\O$, the first order is transformed as
\begin{equation}
\lambda_{(1)}\ri\lambda'_{(1)}=\f{\lambda_{(1)}}{\o^2}-\f{1}{\o^3}\bar{m}^a\bar{m}^b\na_a\na_b\o-\f{2}{\o^4}(\bar{m}^a\na_a\o)^2.\label{eq:lamlam}
\end{equation}
From now on, $'$ represents the quantity after transforming from affine gauge into the gauge such that $\sc$ become compact.

\subsection{Null Generator Congruence and Weyl Curvature}
As mentioned before, the topology of the event horizon is expected to be related to the upper end of $\sc$. We discuss this in the context of asymptotic flatness. In an asymptotic flat spacetime, it is guaranteed that there exists a conformal embedding defined in the latter half of section \ref{sec:2}. The asymptotic flatness, however, accepts the further conformal transformation that is indicated by $\omega$ as gauge freedom. Since the gauge transformation can become singular ($\omega$ can become zero or infinity) at the upper end of $\sc$, we should be careful to choose $\omega$. The gauge freedom can makes the upper end of $\sc$ degenerate into a point or take away to infinity in the direction of null generator tangent $n^a$. Indeed, when we take a gauge choice in which $n^a$ becomes affine parameterized, the null generators are complete in affine parameterizing by definition. Nevertheless, since we want to study the geometrical structure near the upper end of $\sc$, it is important to choose a gauge $\Omega'=\omega\Omega$ in which the null generators are incomplete in affine parameterizing though is complete in a original parameterization $n^a=(\partial/\partial u')^a$. This aspect results from the compactness of $\sc$ (or unphysical manifold) in the upper direction. Under this gauge choice, we discuss the topology of the upper end of $\sc$. 

If we allow for $\omega$ to become angular depend irregularity at the upper end of $\sc$, the irregular angular dependence also may change the topology of the upper end of $\sc$. Then we do not accept such the dependence for the gauge transformation $\Omega'=\omega\Omega$ from affine gauge. 

Newman-Penrose equation\cite{Newman Penrose} relates the shear on $\sc$ to Weyl curvature as the following.
\beq
-n^a\na_a\lambda+\bar{m}^a\na_a\nu&=&-(\mu+\bar{\mu})\lambda-(3\gamma-\bar{\gamma})\lambda\no\\
                                  &+&(3\alpha+\bar{\beta}+\pi-\bar{\tau})\nu+\Psi_{(4)}.\label{eq:NP}
\eeq

We expand each term of the above equation in powers of $\O$ and leave the only leading term. From (\ref{eq:lam}) the order of $\lambda$ is
\begin{equation}
\lambda=O(\O),\label{eq:lambda}
\end{equation}
which is independent of the choice of gauge.
The spin connection $\nu$ can be written as
\begin{equation}
\nu=-\f{1}{2}m^a\na_ak_{(1)}\O+O(\O^2)
\end{equation} 
by the equations (\ref{eq:nn}) and (\ref{eq:orthogonal}). We can choose $\o$ satisfying
\begin{equation}
m^a\na_ak_{(1)}=0\label{eq:omega}
\end{equation}
from the relation (\ref{eq:mupk}). 
For we can write $m^a$ as $m^a=m^{\theta}(\pa/\pa\theta)^a+im^{\phi}(\pa/\pa\phi)^a$ using two spacelike vectors, $(\pa/\pa\theta)^a, (\pa/\pa\phi)^a$ orthogonal to $n^a, l^a$. Then if we transform $\omega$ from affine gauge into another gauge, from the equation (\ref{eq:mm}), the equation (\ref{eq:omega}) becomes
\beq
\f{\pa}{\pa\theta}\biggl(\f{n^a\na_a\omega}{\omega^2}\biggr)=\f{\pa}{\pa\phi}\biggl(\f{n^a\na_a\omega}{\omega^2}\biggr)=0.
\eeq
We may choose $\omega$ so that $n^a\na _a\o /\o^2$ does not depend on $\theta,\phi$ and $\o$ is positive everywhere.  
Therefore, we see that for $\o$ satisfying (\ref{eq:omega}), the order of $\nu$ is
\begin{equation}
\nu=O(\O^2).\label{eq:nu}
\end{equation}

Finally, let us compute the order of the spin connection $\gamma$ defined by $\gamma\equiv 1/2(n^an^b\na_al_b-n^a\bar{m}^b\na_al_b) $. The real part of $\gamma$ is
\beq
\gamma+\bar{\gamma}&=&n^an^b\na_al_b\no\\
                   &=&\mu_{(0)}+O(\O)\label{eq:gpg}
\eeq
from the equations (\ref{eq:lnn}) and (\ref{eq:mpk}).  Since the imaginary part of $\gamma$ becomes 
\begin{equation}
\gamma-\bar{\gamma}=-n^a\bar{m}^b\na_am_b,
\end{equation}
it depends on the direction of $m^a$ which we have not yet determined. Now let us determine the direction of $m^a$ as follows.
We determine the direction of $m^a$ so that on $\sc$, it will be parallelly transported along the null geodesic generators on $\sc$. That is,
\begin{equation}
n^a\na_am^b=0\mbox{ on $\sc$.}
\end{equation}   
Moreover in the direction away off $\sc$, it is parallelly transported along $-l^a$, that is,
\begin{equation}
l^a\na_am^b=0\ \mbox{ everywhere.}
\end{equation} 
Thus, $\gamma-\bar{\gamma}$ becomes
\begin{equation}
\gamma-\bar{\gamma}=O(\O).\label{eq:gmg}
\end{equation}
So we see that from the equations (\ref{eq:gpg}) and (\ref{eq:gmg}), the order of $\gamma$ is
\begin{equation}
\gamma=\f{\mu_{(0)}}{2}+O(\O).\label{eq:gamma}
\end{equation}

From (\ref{eq:lambda}), (\ref{eq:nu}) and (\ref{eq:gamma}), we can write the equation (\ref{eq:NP}) as 
\begin{equation}
n^a\na_a\lambda_{(1)}=3\mu_{(0)}\lambda_{(1)}-\Psi_{4(1)}
\end{equation} 
in the leading order. Integrating this, we obtain the relation
\begin{equation}
\lambda_{(1)}=\exp \bl(3\int\mu_{(0)}du'\br)\int^{u'}du'\Psi_{4(1)}\bl(-3\int\mu_{(0)}du'\br),\label{eq:shear}
\end{equation}
where $u'$ is the parameter of null geodesics on $\sc$ and is defined by $n^a\equiv(\pa/\pa u')^a$. If we transform conformal factor from affine gauge into the gauge such that $\sc$ is compact and put $\o=u^{\prime -\alpha}$, $\mu_{(0)}$ is
\begin{equation}
\mu'_{(0)}=\f{\alpha}{u'}.
\end{equation}
So if the behavior of Weyl curvature on $\sc$ is $\Psi_{4(1)}\sim 1/u^{\prime\epsilon}$ as $u'\ri\infty$, the ratio of the shear to the expansion is 
\begin{equation}
\biggl|\f{\lambda'}{\mu'}\biggr|\sim u^{\prime 2-\epsilon}\O'+O(\O^{\prime 2}),\label{eq:ratio}
\end{equation}
where $\epsilon$ depends on $\omega$ and is a function of $\alpha$. 
Therefore, if there is $\alpha$ satisfying $2-\epsilon>0$, this ratio can diverge, that is, the shear can be much larger than the expansion. This implies that $\sc$ pinches at the upper end. The point is that if  Weyl curvature falls off later than some power of parameter of null generators, two black holes will not coalesce far in the future.

Now that we obtain the sufficient preparation, we can restate our sufficient condition for not coalescing black holes as follows;

\newtheorem{con}{Condition That Black Holes Do Not Coalesce}
\renewcommand{\thecon}{}

\begin{con}
When we transform the conformal factor from affine gauge into another gauge $\O\ri\O'=\O\o$, the topology of future null infinity pinches at the upper end if there exists an unphysical spacetime $(\ti{M},\o^2\ti{g}_{ab})$ with the conformal factor $\o$ satisfying the below conditions.\\
(1)There is a real number $\alpha$ such that $\o=\o(\theta,\phi)u^{\prime -\alpha}$ and $\lim_{u\ri\infty}\o=0$, where $u$ and $u'$ are parameters of the null geodesics on $\sc$ in affine gauge and in another gauge, respectively and $\o(\theta.\phi)$ is the smooth function on a sphere which is positive and not singular everywhere.\\
(2)There is a real number $\alpha>0$ satisfying the following condition;
For any number $L>0$, there are positive numbers $K$ and $\delta$ such that if $u'>K, 0<\Omega'<\delta$, then $|\Omega'u^{\prime 2-\epsilon}|>L$, where $\epsilon$ is an exponent appearing in $\Psi_{4(1)}$ extended in the power of $u'$, that is, $\Psi_{4(1)}\sim u^{\prime -\epsilon}$.
\end{con}

Of course, in such an unphysical spacetime, the spacetime is confomally embedded as illustrated in FIG.1(b).

\section{\label{sec:5}Application to C-metric }
In this section we demonstrate the above condition by an example. Now we consider the vacuum C-metric as the example of the spacetime where two black holes does not eternally coalesce \cite{Cornish and Uttley},\cite{Bonnor},\cite{Bonnor2}. 
The line element of the vacuum C-metric is written in the form of
\beq
ds^2&=&-Hdu^2+H^{-1}dr^2+2Ar^2H^{-1}drdx\no\\
    &{}&+(G^{-1}+A^2r^2H^{-1})r^2dx^2+r^2Gd\varphi^2,\label{eq:cmetric}
\eeq
where 
\beq
H&=&1-2mr^{-1}+6Amx+ArG_{,x}-A^2r^2G,\\
G&=&1-x^2-2Amx^3.
\eeq
BMS coordinate\cite{BMS} is expanded at large distance, but 
since in our discussion it is sufficient to consider the only neighborhood of $\sc$, we may transform the metric (\ref{eq:cmetric}) to a BMS coordinate written by
\beq
ds^2&=&-fdu^2-2e^{2\beta}dudr-2Udud\theta\no\\
    &{}&+r^2(e^{2\gamma}d\theta^2+e^{-2\gamma}\sin^2\theta d\phi^2),\label{eq:BMS}
\eeq
where four functions are 
\beq
f(u,r,\theta)      &=& 1-\f{2M(u,\theta)}{r}+O\bl(\f{1}{r^2}\br),\label{eq:f}\\
\beta(u,r,\theta)  &=& -\f{c^2(u,\theta)}{4r^2}+O\bl(\f{1}{r^3}\br),\label{eq:b}\\
\gamma(u,r,\theta) &=& \f{c(u,\theta)}{r}+O\bl(\f{1}{r^2}\br),\label{eq:g}\\
U(u,r,\theta)      &=& -(c_{\theta}+2c\cot\theta)+O\bl(\f{1}{r}\br),\label{eq:U}
\eeq
and $M(u,\theta)$ is given by
\begin{equation}
M_{,u}=-c_{,u}^2+\f{1}{2}(c_{,\theta\theta}+3c_{,\theta}\cot\theta -2c)_{,u}.
\end{equation}
We should note that these four functions are expressed by one function $c(u,\theta)$ whose differentiation by $u$ is called \textit{the news function}.  
We introduce the new coordinate by
\beq
v\equiv u+2r.
\eeq
In this coordinate, the metric (\ref{eq:BMS}) becomes the following form.
\beq
ds^2&=&-(f-e^{2\beta})du^2-e^{2\beta}dudv-2Udud\theta\no\\
    &{}&+\f{1}{4}(v-u)^2(e^{2\gamma}d\theta^2+e^{-2\gamma}\sin^2\theta d\phi^2). 
\eeq
Furthermore, we introduce the new coordinate $V=1/v$ so that infinity along outgoing null geodesics will correspond to $V=0$. The metric components in the new coordinate $(u,V,\theta,\phi)$ are
\beq
ds^2&=&-(f-e^{2\beta})du^2+\f{e^{2\beta}}{V^2}dudV-2Udud\theta\no\\
    &{}&+\f{1}{4}(v-u)^2(e^{2\gamma}d\theta^2+e^{-2\gamma}\sin^2d\phi^2). 
\eeq 
In turn, we transform the above physical metric.
\beq
d\ti{s}^2&=&-(f-e^{2\beta})\O^2du^2+e^{2\beta}dud\O-2U\O^2dud\theta\no\\
         &{}&+\f{1}{4}(1-\O u)^2(e^{2\gamma}d\theta^2+e^{-2\gamma}\sin^2\theta d\phi^2),\label{eq:conformalBMS}
\eeq
where the unphysical metric is related to the physical metric by $\ti{g}_{ab}=\O^2g_{ab}$ and $\O=V$.
    
Four functions (\ref{eq:f}),(\ref{eq:b}),(\ref{eq:g}) and (\ref{eq:U}) are expanded as
\beq
\beta &=& -c^2(u,\theta)\O^2+O(\O^3),\\
\gamma&=& 2c(u,\theta)\O+O(\O^2),\\
f     &=& 1-4M(u,\theta)\O+O(\O^2),\\
U     &=& -(c_{,\theta}+2c\cot\theta)+[-2c(c_{,\theta}+2c\cot\theta)\no\\
      &{}&+2(2c_{,u}+3cc_{,\theta}+4c^4\cot\theta)]\O+O(\O^2)
\eeq
in powers of conformal factor $\O$. 

Now let us determine the basis vector mentioned in section \ref{sec:4} from the metric section (\ref{eq:conformalBMS}). It is easy to decompose the metric (\ref{eq:conformalBMS}) into the form of
\begin{equation}
\ti{g}_{ab}=(k_{(1)}\O+\cdots)l_al_b-2l_{(a}n_{b)}+2m_{(a}\bar{m}_{b)}.
\end{equation}
We see that the four dual vectors are 
\beq
n_a&=&(d\O)_a,\label{eq:defn}\\
l_a&=&-\f{1}{2}e^{2\beta}(du)_a,\\
m_a&=&\f{1}{2\sq{2}}(1-u\O)\bl(-e^{-\gamma}\f{4U\O^2}{(1-u\O)^2}(du)_a\no\\
   &{}&+e^{\gamma}(d\theta)_a-ie^{-\gamma}\sin\theta(d\phi)_a \br),\\
\bar{m}_a&=& \f{1}{2\sqrt{2}}(1-u\O)\bl(-e^{-\gamma}\f{4U\O^2}{(1-u\O)^2}(du)_a\no\\
   &{}&+e^{\gamma}(d\theta)_a+ie^{-\gamma}\sin\theta(d\phi)_a \br),  
\eeq
and we raise the indices of the dual vectors to obtain four basis vectors.
\beq
n^a &=& 2e^{-2\beta}\bl(\f{\pa}{\pa u}\br)^a\no\\
    &+& [4(f-e^{2\beta})e^{-4\beta}\O^2+16e^{-4\beta-2\gamma}\f{U^2\O^4}{(1-u\O)^2}]\bl(\f{\pa}{\pa \O}\br)^a\no\\
    &+& 8e^{-2\gamma-2\beta}U\f{\O^2}{(1-u\O)^2}\bl(\f{\pa}{\pa \theta}\br)^a,\label{eq:n}\\
l^a&=&-\bl(\f{\pa}{\pa \O}\br)^a,\\
m^a &=&\f{\sq{2}e^{-\gamma}}{1-u\O}\bl(\f{\pa}{\pa \theta}\br)^a+i\f{\sq{2}e^{\gamma}}{\sin\theta(1-u\O)}\bl(\f{\pa}{\pa \phi}\br)^a,\\
\bar{m}^a &=&\f{\sq{2}e^{-\gamma}}{1-u\O}\bl(\f{\pa}{\pa \theta}\br)^a-i\f{\sq{2}e^{\gamma}}{\sin\theta(1-u\O)}\bl(\f{\pa}{\pa \phi}\br)^a.
\eeq
From (\ref{eq:defn}) and (\ref{eq:n}), we can compute the norm of $n^a$ as follows.
\begin{equation}
n^an_a=4(f-e^{2\beta})e^{-4\beta}\O^2+16e^{-4\beta-2\gamma}\f{U^2\O^4}{(1-u\O)^2}.\label{eq:nn3}
\end{equation}
Since the first order in the equation (\ref{eq:nn3}) vanishes, we see that the choice of conformal factor $\O=V$ is affine gauge. That is, in the choice of this gauge
\begin{equation}
\mu_{(0)}=-p_{(0)}=-\f{1}{2}k_{(0)}=0
\end{equation}
holds. The shear $\lambda$ in this gauge is 
\beq
\lambda&\equiv& -\bar{m}^a\bar{m}^b\na_an_b\no\\
       &=&     -4c_{,u}\O+O(\O^2)\\
       &=&     \f{3a^3m}{A^3}\f{\sin^2\theta}{u^4}\O+O(\O^2),
\eeq
where we used the fact the news function of the vacuum C-metric~\cite{Tomimatsu} is
\begin{equation}
c_{,u}=\f{-3a^3m}{4A^3}\f{\sin^2\theta}{u^4},
\end{equation}
and $u$ is the affine parameter of the null generator of $\sc$. 

We must transform conformal factor from affine gauge into the gauge such that $\sc$ is compact and examine the ratio of the shear to the expansion of the null geodesic generators which plunge into the zero at the upper end of $\sc$. 
In affine gauge, the shear in this direction behaves as $u\ri\infty$ and $\Omega\ri 0$
\begin{equation}
\lambda\sim\lambda_{(1)}\Omega\sim\f{1}{u^4}\O\label{eq:lamu}
\end{equation} 
in affine gauge. Under the transformation $\O\ri\O'=\o\O$, there is the relation between affine parameter and the parameter of the geodesic in new gauge as follows
\beq
u &=& \left\{
             \begin{array}{ll}
               \f{u^{\prime \alpha+1}}{\alpha+1} & \mbox{@($\alpha\ne -1$)}\\
               \log u'             & \mbox{@($\alpha = -1$),}\\
              \end{array}\label{eq:uu}
       \right.       
\eeq

\noindent
where we set $\o=1/u^{\prime\alpha}$.

We can obtain the ratio of the shear to the expansion from (\ref{eq:shear}).
\begin{equation}
\biggl|\f{\lambda'}{\mu'}\biggr|\sim u^{\prime\alpha+1}\O',
\end{equation}
where we used (\ref{eq:mm}), (\ref{eq:lamlam}), (\ref{eq:lamu}) and (\ref{eq:uu}).
So, we see that when $u\ri\infty (u'\ri\infty)$, this ratio diverges to infinity. We ensured the existence of the comformal factor $\O'=\o\O$ such that the ratio of the shear to the expansion of the null generators on $\sc$ diverges. We also find easily 
\begin{equation}
\lim_{u\ri\infty}\o=0.
\end{equation}
On the other hand, though Newman-Penrose formalism is ,of course, true, we confirm that the behavior of Weyl curvature is obtained directly from the metric
\beq
\tilde{\Psi}'_{4(1)}\sim\f{1}{u^{\prime 1-\alpha}}.
\eeq 

So if we choose $\omega$ so that $0<\alpha<1$, $0<\epsilon<2$ holds. 
Therefore, two conditions (1) and (2) stated in the end of section \ref{sec:4} are satisfied. 
This implies that cross section of $\sc$ pinches at the upper end, as we have expected.

\section{\label{sec:8}Summary and Discussion}
We have investigated the relation between the topological structure of an event horizon far in the future and the geometrical structure of the future null infinity. A sufficient condition for topologically non-trivial final black holes is proposed and applied to C-metric as an example. If some component of Weyl curvature has a late power law tail, since future null infinity pinches off at the upper end, two black holes will not coalesce far in the future. We make sure of the validity of our method on C-metric. 

While we concentrate on the gravitational radiation $\Psi_{(4)}$ of Weyl curvature as leading order contribution, there are possibilities that other component of curvature may contribute to the topological structure of future null infinity, for example, for Majumdar-Papapetrou solution. In Majumdar-Papapetrou spacetime, since gravitational field radiation $\Psi_{(4)}$ vanishes, we expect $\Psi_{(2)}$ contributes to the topological structure of the upper end of $\sc$ in a higher order.    

In the present article, we mainly considered the case with several spherical black holes, since Wald and Chru\'sciel proved in a stationary spacetime the black holes are spherical under null energy condition. Therefore, if we expect the final state of a black hole spacetime become stationary, our method will give the number of black hole black holes with a spherical topology far in the future. 

Using the result of this article, what can we know about the collapsing stars from the observation of the gravitational radiation. The sufficient condition for topologically non-trivial black hole suggests the power of late time tail radiation will let us know about the final topology of black holes. Of course, to get rigorous statement about that, we should make clear the astrophysical meaning of the gauge choice used in this article. 

\begin{acknowledgments}
We thank Dr. Nakamura for useful comments.
We would like to thank Professor A. Hosoya and T. Shiromizu for continuous encouragement.
\end{acknowledgments}

\end{document}